\documentclass[aps,prl,twocolumn,groupedaddress]{revtex4}
\usepackage{graphicx}

\begin{document}

\title{Angle integrated photoemission study of SmO$_{0.85}$F$_{0.15}$FeAs}

\author{H. W. Ou$^{1}$, J. F. Zhao$^{1}$, Y. Zhang$^{1}$, D. W. Shen$^{1}$, B. Zhou$^{1}$,
L. X. Yang$^{1}$, C. He$^{1}$, F. Chen$^{1}$, M. Xu$^{1}$, T.
Wu$^2$, X. H. Chen$^2$, Y. Chen$^{1}$, and D. L. Feng$^{1}$}
\email{dlfeng@fudan.edu.cn}

\affiliation{$^1$Department of Physics, Surface Physics Laboratory
(National Key Laboratory), and Advanced Materials Laboratory, Fudan
University, Shanghai 200433, P. R. China}

\affiliation{$^2$Hefei National Laboratory for Physical Sciences at
Microscale and Department of Physics, University of Science and
Technology of China, Hefei, Anhui 230026, P. R. China}

\date{\today}

\begin{abstract}
The electronic structure of the new superconductor,
SmO$_{1-x}$F$_x$FeAs ($x=0.15$), has been studied by
angle-integrated photoemission spectroscopy. Our data show a sharp
feature very close to the Fermi energy, and a relative flat
distribution of the density of states between 0.5\,eV and 3\,eV
binding energy, which agrees best with band structure calculations
considering an antiferromagnetic ground state. No noticeable gap
opening was observed at 12 Kelvin below the superconducting
transition temperature, indicating the existence of large ungapped
regions in the Brillouin zone.
\end{abstract}

\maketitle

The discovery of superconductivity with an unexpectedly high
superconducting transition temperature ($T_c$) of 26K in the
iron-based LaO$_{1-x}$F$_x$FeAs ($x=0.05-0.12$) \cite{JACS} has
ignited intensive studies
recently\cite{LDA0,LDA1,LDA2,LDA3,LDA4,DMFT,Phonon,XDai,NLWang1,HHWen1,HHWen2,NLWang2,ChenXH,Chen}.
Particularly, Chen \textit{et al.} has raised the $T_{c}$ to  43\,K
with SmO$_{1-x}$F$_x$FeAs ($x=0.15$)\cite{ChenXH}, a record high for
non-cuprate superconductors. Such a high $T_c$ is hard to understand
within the conventional BCS mechanism for
superconductivity\cite{McMillan}. Up to now, multiple pieces of
evidences have been gathered in recent studies to unveil the nature
of these novel superconductors. Experiments from the specific heat
measurements~\cite{HHWen1}, point-contact tunneling spectroscopy
~\cite{HHWen2},  and infrared reflectance
spectroscopy~\cite{NLWang2} provided support of the existence of
unconventional superconductivity in such materials.

The band structure of LaO$_{1-x}$F$_x$FeAs has been calculated by
density functional theory (DFT) and dynamical mean field theory
(DMFT), based on which various theoretical proposals have been put
forward. For example, it has been revealed that the electron-phonon
interaction in the system is too small to support such high
$T_c$'s~\cite{Phonon},  while various unconventional superconducting
pairing symmetries have been proposed\cite{LDA2,XDai,Chen}.
Considering the possible strong correlation involved, these results
remain to be verified experimentally. However, besides a few pieces
of transport measurements on polycrystalline samples, no direct
measurement of the electronic structure has been reported so far.

In this paper, we report angle integrated photoemission spectroscopy
measurement of a  SmO$_{0.85}$F$_{0.15}$FeAs polycrystalline sample,
which gives the density of states (DOS) of the system. A relatively
narrow feature at 0.25\,eV was observed, which agrees very well with
DFT calculations based on an antiferromagnetic ground state. No
noticeable gap opening was observed even at 12 Kelvin below the
superconducting transition temperature, indicating the existence of
large ungapped regions in the Brillouin zone. Our data put strong
constraints on theoretical studies.

\begin{figure}[b!]
\includegraphics[width=8.5cm]{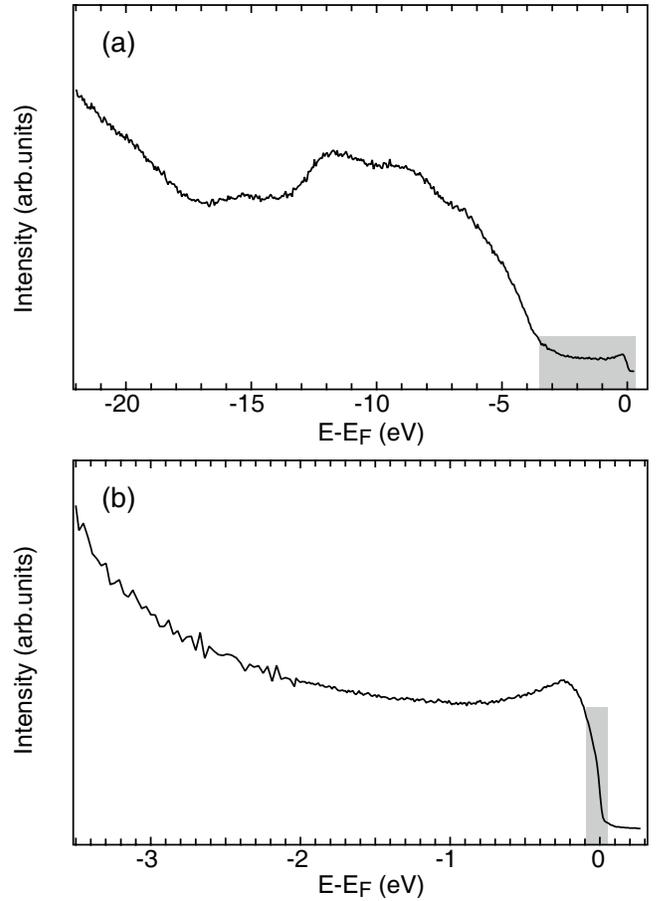}
\caption{Angle integrated spectrum of  SmO$_{1-x}$F$_x$FeAs (a) over
a large energy window, and (b) within the shaded region of (a). Data
in (a) and (b) were taken at 45K with 40.8eV and 21.2eV photons
correspondingly. }
\end{figure}

SmO$_{0.85}$F$_{0.15}$FeAs polycrystal has been synthesized through
solid state reaction, which enters the superconducting state at
43\,K. The detailed synthesis and characterization of the sample has
been described elsewhere\cite{ChenXH}. Photoemission measurements
were performed with Helium-I $\alpha$(21.2\,eV) and
Helium-II(40.8\,eV) emission lines of a Helium discharge lamp. Data
were taken with a Scienta R4000 electron analyzer.  The energy
resolution is 10\,meV. The sample rod was broken \textit{in-situ
}and then measured in ultra-high vacuum ($\sim 3\times
10^{-11}\,mbar$). The isotropic distribution of the polycrystal
orientation was confirmed by varying the photoelectron emission
angle.

The angle integrated spectrum of  SmO$_{0.85}$F$_{0.15}$FeAs shown
in Fig.\,1(a-b) measures the DOS of the system. Except the low
energy feature near the Fermi energy ($E_F$), it is quite flat
within the first 3\,eV below $E_F$, and stronger features only show
up from 3\,eV binding energy and higher. The low energy feature
was attributed to Fe $3d$ states in various DFT band structure
calculations\cite{LDA0,LDA1,LDA2,LDA3,LDA4,note}. However, there are
quantitative differences amongst them. Our data, especially the peak
at 0.25\,eV and the flat distribution between 0.5\,eV to 3\,eV
binding energy, agree best with the DFT calculation that considers
an antiferromagnetic ground state\cite{LDA3,LDA4}. On the other
hand, this low energy feature does not show up in dynamical mean
field theory calculation~\cite{DMFT}.

\begin{figure}[t]
\includegraphics[width=8.5cm]{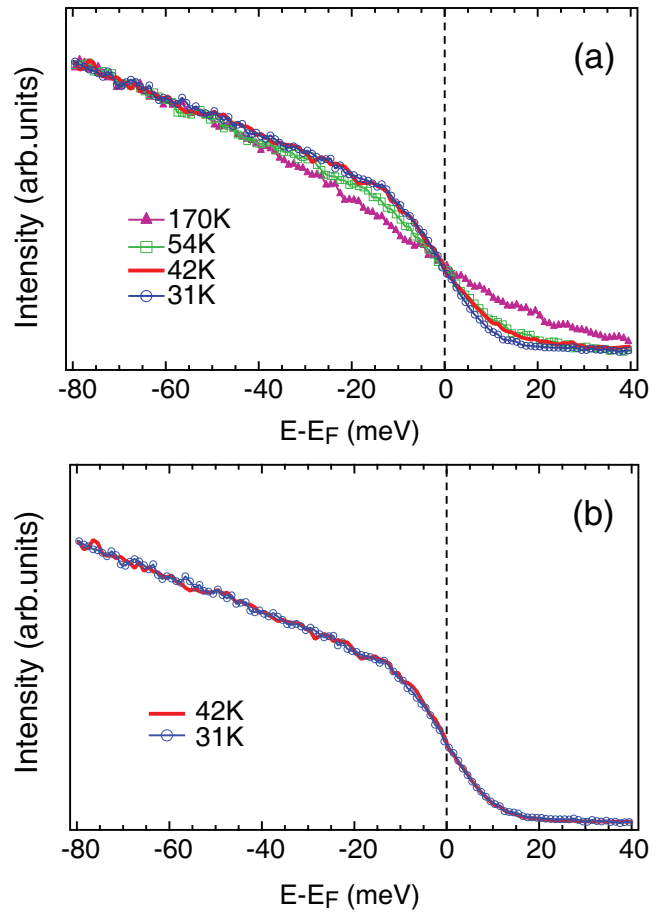}
\caption{(color online). (a) Temperature dependence of the spectrum
in the shaded energy window of Fig.1(b). Data were taken with 21.2eV
photons. (b) compares the spectra at 31K and 42K, where the 42K
spectrum has been divided by resolution convoluted Fermi function at
42K and then multiplied with that at 31K. }
\end{figure}

In order to study the superconducting gap, the angle-integrated
spectrum near $E_F$ is taken as a function of temperature (Fig.2a).
There is a clear cutoff by the Fermi function, and the middle point
of the leading edge crosses the Fermi energy for all spectra. Fig.2b
compares the 31K spectrum with 42K spectrum after removing the
thermal broadening effects due to their temperature difference
(details in caption). It seems they overlap very well. For an s-wave
superconductor, at a temperature of 75\% $T_c$, more than 60\% of
the gap would have already opened. The absence of \textit{any} sign
of gap-opening  in the DOS measured here might be partially
attributed to the broad transition in this polycrystalline sample.
Nevertheless it does indicate that large portions of the Fermi
surface are still ungapped at 12\,K below $T_c$ for
SmO$_{0.85}$F$_{0.15}$FeAs, which is a sign of unconventional
superconductivity. Future measurements on high quality single
crystal are necessary to further clarify this issue. Moreover, we
note that an anomaly at about 150K was observed in the resistivity
data\cite{ChenXH}, which resembles the pseudogap effects in cuprate
superconductors. However, our data taken above and below 150K did
not show any sign of pseudogap opening.


To summarize, we have reported angle-integrated photoemission
spectroscopy results of the new superconductor
SmO$_{0.85}$F$_{0.15}$FeAs. Our data, particularly the sharp feature
at 0.25\,eV below the Fermi energy, put quite strong constraints on
various theoretical calculations. No appreciable gap effects on the
density of states were observed at 12 Kelvin below the
superconducting transition temperature, indicating possible
existence of large ungapped regions in the Brillouin zone.

We gratefully acknowledge the helpful discussions with Prof. S. Y.
Li. This work was supported by the Nature Science Foundation of
China and by the Ministry of Science and Technology of China
(National Basic Research Program No.2006CB921300 and 2006CB922005),
and STCSM of China.


\begin{references}
\bibitem{note}We note that although Sm
replaces La here, they both would contribute negligibly to the
shallow band structure. Therefore, one can compare our data with the
band structure calculation of LaO$_{1-x}$F$_x$FeAs.
\end{references}
\end{document}